\pdfoutput=1

\documentclass[11pt,letterpaper]{article}
\usepackage{jheppub}
\usepackage[all]{xy}
\usepackage{xcolor}
\usepackage{url}
\def\ket#1{\langle #1 \rangle}

\DeclareMathOperator{\Conf}{Conf}
\DeclareMathOperator{\Gr}{Gr}
\DeclareMathOperator{\Li}{Li}

\preprint{Brown-HET-1656}

\title{An analytic result for the two-loop seven-point MHV amplitude in $\mathcal{N}=4$ SYM}

\author{John Golden,}
\author{Marcus Spradlin}

\affiliation{Department of Physics, Brown University\\
Box 1843\\
Providence, RI 02912-1843,
USA}

\abstract{
We describe a general algorithm which builds
on several pieces of data available in the literature
to construct explicit analytic formulas
for two-loop MHV amplitudes in $\mathcal{N}=4$ super-Yang-Mills theory.
The non-classical part of an amplitude is built from $A_3$ cluster
polylogarithm functions; classical polylogarithms with (negative)
cluster $\mathcal{X}$-coordinate arguments are added to complete the
symbol of the amplitude; beyond-the-symbol terms proportional to
$\pi^2$ are determined by comparison with the differential of the amplitude;
and the overall additive constant is fixed by the collinear limit.
We present an explicit formula for the seven-point amplitude
$R_7^{(2)}$ as a sample application.
}

\begin{document}
\maketitle

\section{Introduction}

This note is a natural continuation of the research
program that has been pursued
in the
papers~\cite{Goncharov:2010jf,Golden:2013xva,Golden:2013lha,Golden:2014xqa}
and has been heavily guided by earlier mathematical work of
Goncharov on both the structure of polylogarithm functions and on
cluster algebras (see in particular~\cite{G02} and~\cite{FG03b}).
The physics
goal of our program is, narrowly, to understand the rich mathematical
structure of two-loop amplitudes in $\mathcal{N}=4$ supersymmetric
Yang-Mills (SYM) theory~\cite{Brink:1976bc},
and more broadly, to develop a toolkit of
mathematical
techniques useful for unlocking the structure of
multi-loop amplitudes
in general field theories.
An example of the latter
is the symbol calculus, which following~\cite{Goncharov:2010jf}
has become a very useful workhorse for dealing with the kinds of polylogarithm
functions which are ubiquitous in multi-loop calculations, while the intimate
connection between
amplitudes and cluster algebras
unearthed in~\cite{Golden:2013lha}
is a prime example of the very special
structure exemplified by SYM theory in particular.

In this paper we tie together several threads which have run
through the earlier
work~\cite{Goncharov:2010jf,Golden:2013xva,Golden:2013lha,Golden:2014xqa}
but have not yet been fully wrapped up.
Our immediate goal will be to construct
an explicit analytic formula for the two-loop seven-point
MHV amplitude $R_7^{(2)}$ in SYM theory\footnote{More precisely
$R_n^{(L)}$ stands for the $n$-particle $L$-loop
{\it remainder function}, after the infrared singularities of the
amplitude have been subtracted in a now standard way
following~\cite{Anastasiou:2003kj,Bern:2005iz}.
Dual conformal
symmetry requires $R_n^{(L)}$ to vanish for $n<6$
at any loop order~\cite{Drummond:2007cf,Drummond:2007au},
but a numerical study~\cite{Drummond:2007bm}
established that $R_6^{(2)}$ is nonzero.}.
While it may be interesting in its own
right, we do not view the formula itself as the primary result of this
paper.  Rather our aim is to first review the various obstacles that
arise in the pursuit of writing such analytic formulas, and then to bring
together
the relevant ideas and results
from~\cite{Goncharov:2010jf,Golden:2013xva,Golden:2013lha,Golden:2014xqa,futureWork}
to argue that the problem of constructing analytic formulas for
$R_n^{(2)}$ for any desired $n$ may be considered ``solved'' (modulo the availability
of sufficient computer power, of course).
By this we mean that we describe an algorithm which,
building on the scaffolding provided by
Caron-Huot's
computation~\cite{CaronHuot:2011ky}
of the symbol of $R_n^{(2)}$, may be used to construct an analytic
formula for any desired $n$.
The result for $n=6$ appeared in~\cite{Goncharov:2010jf},
and
we present a result for $n=7$ here as a specific application of our algorithm.
Numerical studies of $R_n^{(2)}$ have been carried out
for $n=6$
in~\cite{Drummond:2007bm,Bern:2008ap,Drummond:2008aq} and for higher $n$
in~\cite{Anastasiou:2009kna,Brandhuber:2009da}, and
explicit formulas are known for the special case when
all particles have momenta lying in a common $\mathbb{R}^{1,1}$
subspace
of four-dimensional Minkowski space~\cite{DelDuca:2010zp,Heslop:2010kq,Gaiotto:2010fk}.

We do not address here the question of how the computational complexity of our
algorithm scales with $n$ because we hope that this will ultimately
be an irrelevant question.  As has happened often before in physics,
and especially so in the study of SYM theory, we believe that
once suitably packaged and digestible results accumulate for various
relatively small values of $n$,
the structure might become clear enough that one can extrapolate
an all-$n$ formula, which could subsequently be proven to be correct or
at least could be checked to be consistent with all known properties of
the true
amplitudes.

Amplitudeology is a data-driven enterprise where insights gleaned
by analyzing the results of a seemingly difficult calculation have often
revealed hidden structure which trivialize the original calculation,
and help to make the next set of calculations simpler (or even just
possible).  We very much anticipate that the formula we obtain for
$R_7^{(2)}$ will not be the simplest or ``best'' one possible,
but hope that the algorithm
described in this paper will prove useful for generating new
data for the amplitude community.

Section 2 contains some  brief background material and definitions.
Section 3 comments on the difficulties of
integrating symbols in general, and on the tools we employ to overcome
these difficulties.  We also discuss the relation of our work to
a complementary approach to similar problems which has been used
by Dixon and collaborators to achieve several impressive results on
multi-loop six-point amplitudes~\cite{Dixon:2011nj,Dixon:2011pw,Dixon:2013eka,Dixon:2014voa}.
Section 4 outlines our general algorithm, while section 5 discusses its
application to the specific case of $R_7^{(2)}$, culminating in the construction
of a complete analytic formula for this amplitude, some properties of
which are discussed in section 6.

\section{Background}

This section is a brief review of some of the more advanced mathematics that will appear throughout the rest of the paper, namely the coproduct $\delta$ and cluster algebras. For a more thorough introduction to these topics, see~\cite{Golden:2013xva,Golden:2014xqa}.

The space of polylogarithm functions modulo products is a Lie coalgebra with
coproduct\footnote{Throughout this paper, we use the word ``coproduct''
to denote $\delta$, which satisfies $\delta^2 = 0$, rather than $\Delta$ which operates
by simply deconcatenating the symbol.  We refer the
reader to~\cite{Duhr:2011zq,Golden:2013xva}
for additional details.} $\delta$.
The coproduct maps a polylogarithm function of weight 4 (the case of relevance
to two-loop amplitudes) into two component spaces,
$\Lambda^2 B_2$ and $B_3 \otimes \mathbb{C}^*$. Here, $B_k$ refers to the Bloch group, which roughly speaking represents the space of classical weight $k$ polylogarithm functions modulo functional relationships amongst $\Li_k$ and modulo products of functions of lower weight. Elements of $B_k$ are linear combinations
of objects denoted by $\{x\}_k$, which stands for the equivalence class
containing the function $-\Li_k(-x)$.
The $\Lambda^2 B_2$ component of the coproduct captures the obstruction
to writing a function in terms of the
classical polylogarithm functions $\Li_k$~\cite{G91a,G91b}. The $B_3\otimes\mathbb{C}^*$ component of the coproduct encapsulates all of the intrinsically weight 4 terms in a function.

Cluster algebras are generated by a preferred set of variables (``cluster coordinates") grouped in disjoint sets called clusters related to each other by a transformation called mutation. The cluster algebra relevant for two-loop MHV scattering amplitudes in SYM theory is the $\Gr(4,n)$ Grassmannian cluster algebra, which is related to the kinematic configuration space for $n$ particles, $\Conf_n(\mathbb{P}^3)$. These coordinates come in two flavors, $\mathcal{A}$- and $\mathcal{X}$- coordinates. An example of $\mathcal{A}$-coordinates are the standard Pl\"ucker coordinates $\ket{ijkl}= \det(Z_iZ_jZ_kZ_l)$ (in terms of momentum-twistor variables~\cite{Hodges:2009hk}). Slightly more complicated examples that will appear later in this paper are of the type
\begin{align}
\label{eq:aexample1}
\ket{a(bc)(de)(fg)} &\equiv \ket{abde}\ket{acfg} - \ket{abfg} \ket{acde},\\
\label{eq:aexample2}
\ket{ab(cde) \cap (fgh)} &\equiv \ket{acde}\ket{bfgh} -
\ket{bcde}\ket{afgh}.
\end{align}
Cluster $\mathcal{X}$-coordinates are a special class of cross-ratios built from $\mathcal{A}$-coordinates.

These two topics, polylogarithms and cluster algebras, merge beautifuly in the arena of SYM theory. Firstly, only cluster $\mathcal{A}$-coordinates for $\Gr(4,n)$ appear in the symbol for $R^{(2)}_n$. Moreover, the coproduct of $R^{(2)}_7$ was calculated in~\cite{Golden:2013xva} and it was noted that the elements $\{x\}_2$ and $\{x\}_3$ appearing in the coproduct were cluster $\mathcal{X}$-coordinates of the $\Gr(4,7)$ Grassmannian cluster
algebra. Furthermore, it was noted that the function for $R^{(2)}_6$ obtained in~\cite{Goncharov:2010jf} can be written purely in terms of classical polylogarithms $\Li_k$ with (negative) $\mathcal{X}$-coordinates as arguments. In this paper we extend these connections to a general algorithm for constructing the function $R^{(2)}_n$.

Let us note that the $\Gr(4,n)$ cluster algebra has infinitely many
$\mathcal{A}$- and $\mathcal{X}$-coordinates when $n>7$, but we believe
that this presents
no obstruction to our algorithm since it
is evident from the result of~\cite{CaronHuot:2011ky} that
only finitely many (in fact,
precisely $\frac{3}{2}n(n-5)^2$) of the
$\mathcal{A}$-coordinates actually appear in the two-loop MHV amplitude
$R_n^{(2)}$, and our experience has shown that the ``most complicated
part'' of these amplitudes (see~\cite{Golden:2014xqa} for details)
can be expressed in terms of the $\mathcal{X}$-coordinates
belonging to finitely many $A_3$ subalgebras of $\Gr(4,n)$.
For the special cases $n=6,7$, we expect that the two-loop symbol alphabet
(which contains already all available $\mathcal{A}$-coordinates)
will be sufficient to express all amplitudes (whether MHV or not) to all
loop order, but for $n>7$ we know of no reason to exclude the possibility
that the symbol alphabet could grow larger at higher loops (indeed
we expect it to become infinite for ten-point
N${}^3$MHV amplitudes starting already at only two loops).

A salient feature of cluster $\mathcal{X}$-coordinates is that they are positive when evaluated inside the positive Grasmmannian, defined as the subset of the Euclidean domain where $\ket{ijkl}>0$ whenever $i<j<k<l$. This is incredibly important because it allows us to impose analyticity inside the positive domain with relative ease (since $\Li_k(x)$ is smooth for $x<0$), in particular without
having to worry about branch cuts.
It would be interesting to check the extension of our final formula
to more general Euclidean kinematics, for which it
would be necessary to specify where to take the branch cut of each $\Li_k(x)$
(as was done for example in~\cite{Goncharov:2010jf} for $n=6$).
It would also be interesting to explore the analytic continuation to other regions
outside the Euclidean domain, for example to make contact
with work on the seven-point amplitude in the multi-Regge
regime~\cite{Bartels:2011ge,Bartels:2013jna,Bartels:2014ppa}.

Before we describe our algorithm we would first like to clarify the difficulties that our cluster algebraic approach allows us to overcome.

\section{The Problem of Integrating Symbols}

The problem of finding an explicit polylogarithm function
whose symbol matches a given random (but integrable) symbol is hopeless; no
algorithm exists in general.  Fortunately, amplitudes in SYM theory
do not have random symbols, nor do we expect them to be expressed in terms
of completely random functions.

In such happier cases the problem can be tractable if
the desired function may be
expressed in terms of some class of generalized polylogarithm functions whose arguments
are all drawn from some particular finite collection of well-behaved variables.
Then the problem of integrating the symbol becomes simply one of linear
algebra:  one writes a general linear combination of the functions in
the ansatz, and chooses the coefficients to match the desired symbol.
Ideally, the ansatz should be just big enough to contain the
answer, and not too big.
If the ansatz is too overcomplete\footnote{Some overcompleteness is
inevitable in our approach
due to $\Li_k$ identities involving configurations of points in
projective space (see for example~\cite{G91a,G91b,G95}), but such identities
are rare when
the arguments are restricted to be (negative) cluster $\mathcal{X}$-coordinates.
The only currently known
non-trivial identities of this type are the
5-term $A_2$ identity (Abel's identity)
for $\Li_2$ and the 40-term $D_4$ identity
for $\Li_3$~\cite{Golden:2013xva}.}
there can be considerable ambiguity in choosing a functional representative
for the integrated symbol.

If one were merely interested in being able to obtain numerical values
for SYM amplitudes, then such ambiguity would be of little concern.
If the goal however
is to unlock their mathematical structure,
then it is desirable to have functional representations
which manifest, to the extent possible, all of their known properties.  From
this point of view, any ambiguity
in how to write an amplitude is seen as an
inefficiency, a wasted opportunity.

In a series of
papers~\cite{Dixon:2011nj,Dixon:2011pw,Dixon:2013eka,Dixon:2014voa},
Dixon and collaborators have pursued one approach to this problem
by studying
``hexagon functions'',
defined as polylogarithm functions whose symbol can be expressed in terms
of a certain 9-letter alphabet (in our terminology, the alphabet
of $\mathcal{A}$-coordinates for the $\Gr(4,6)$ Grassmannian cluster
algebra) and which have the appropriate analytic structure for
scattering amplitudes (specifically, that they must be analytic everywhere
inside the Euclidean domain, with branch points on the boundary
of the Euclidean domain when $\ket{i\,i{+}1\,j\,j{+}1}=0$
for some $i,j$).
By systematically classifying such hexagon functions through weight eight,
and by using physical input about the near-collinear limit derived from
the Wilson loop OPE approach~\cite{Basso:2010in,Basso:2013vsa,Basso:2013aha,Papathanasiou:2013uoa,Basso:2014koa} and from the multi-Regge limit~\cite{Bartels:2008ce,Bartels:2008sc,Schabinger:2009bb,Lipatov:2010qg,Lipatov:2010ad,Bartels:2010tx,Dixon:2011pw,Fadin:2011we,Dixon:2012yy}, they have
determined analytic expressions for the
six-point NMHV amplitude at two loops,
and the six-point MHV amplitude at three and four loops.

It would be extremely interesting to pursue a similar approach for $n>6$,
by exploring for example the space of ``heptagon functions''.  Our
trepidation to take this route stems from the fact that the required
symbol alphabet grows rapidly with $n$:  as mentioned above,
the symbol alphabet
for $R_n^{(2)}$ has $\frac{3}{2} n (n-5)^2$ entries~\cite{CaronHuot:2011ky},
so the space of
weight-four symbols has dimension\footnote{This
is rather too pessimistic; the analyticity condition cuts this down
by one power of $n$ and the integrability condition
no doubt cuts down by some more powers of $n$.}
$\mathcal{O}(n^{12})$.

We have pursued instead the somewhat orthogonal approach of organizing our
calculations not from left-to-right in the symbol, but rather in
order of decreasing mathematical complexity of the functional constituents.
At weight four, this means that we first
focus our attention on the ``non-classical''
part of the amplitude: the $\Lambda^2 B_2$ component of its
coproduct.
The remaining purely classical pieces
of an amplitude can be systematically computed in order from most to least
complicated
by following the procedure outlined in~\cite{Goncharov:2010jf}.
This approach has the disadvantage of leaving the analytic properties of
amplitudes obscure, while it has the advantage of making some remarkable
mathematical properties---the relation to the cluster structure on
the kinematic domain---manifest.

The very first step in this approach is the one most fraught with peril, as
we now explain.  The $\Lambda^2 B_2$ component of the coproduct of
$R_n^{(2)}$ can be expressed~\cite{Golden:2013xva,futureWork}
as a linear combination of various
$\{x_i\}_2 \wedge \{x_j\}_2$ where the $x$'s are drawn from
the $\mathcal{X}$-coordinates
of the $\Gr(4,n)$ cluster algebra.  Moreover, the $x$'s always appear
together in pairs satisfying
$\{ x_i, x_j \} = 0$
with respect to the natural Poisson structure on the kinematic
domain $\Conf_n(\mathbb{P}^3)$; this implies that each pair
of variables
generates an $A_1 \times A_1$ subalgebra of the $\Gr(4,n)$ cluster algebra.

For several years a guiding aim of this research program,
strongly advocated by Goncharov, has been
that it should be possible to write each amplitude under consideration
as a linear combination of special functions associated with
smaller building blocks (``atoms'').  For example, it is well-known that
the function\footnote{We caution the reader that several variants
of this function exist in the literature, beginning
with~\cite{Goncharov-Galois1}, all of which differ from each
other by the addition of terms proportional to $\Li_4$, or products of
lower-weight $\Li_k$'s.  In fact even in this short paper we will
use a second variant $K_{2,2}$ momentarily.
All of these variants have the same $\Lambda^2 B_2$ coproduct
component.  The particular $L_{2,2}(x,y)$ used here may also be expressed
as $L_{2,2}(x,y) = \frac{1}{2} \Li_{2,2}(x/y,-y) - (x \leftrightarrow y)$
in terms of the $\Li_{2,2}$ function.}
\begin{equation}
L_{2,2}(x,y) = \frac{1}{2} \int_0^1 \frac{dt}{t}
\Li_2(-t x) \Li_1(-t y) - (x \leftrightarrow y)
\end{equation}
has the simple $\Lambda^2 B_2$ coproduct component $\{x\}_2 \wedge \{y\}_2$.
Therefore one might be tempted to construct the non-classical part of
a desired $R_n^{(2)}$ by writing down an appropriate linear combination
of $L_{2,2}(x_i,x_j)$ functions; the difference between this object and
$R_n^{(2)}$ must then be expressible in terms of the classical functions
$\Li_k$ only.

The fatal flaw in this approach is that while $L_{2,2}(x_i,x_j)$ indeed
has a simple
coproduct, it is poorly adapted to applications where one wants
to manifest cluster structure because its symbol has some entries
of the form $x_i - x_j$,
which is never expressible as a product of cluster $\mathcal{A}$-coordinates
(and thus can never be an $\mathcal{X}$-coordinate).
Therefore one would have to considerably enlarge the symbol alphabet
under consideration in order to fit all of the classical pieces of the
amplitude left over by subtracting a linear combination of
$L_{2,2}$'s.  Just as bad, one would almost inevitably generate
$\Li_k$ functions whose arguments range over the entire real line,
greatly complicating the problem of arranging all of the branch cuts
of the individual terms to conspire to cancel out everywhere in the positive domain.

So if we want to maintain a connection to the cluster structure (and,
more practically, to avoid enormously complicating the calculation by
being forced to clean up unwanted mess in the symbol),
we should abandon the idea that each individual term $\{x_i\}_2 \wedge
\{x_j\}_2$ may be thought of as an atom\footnote{Instead they are perhaps
quarks: never allowed to appear alone, but always bound safely together
in $A_2$ functions
or perhaps other, not yet discovered, more exotic baryons.}.
The problem of identifying the smallest building block
manifesting all of the known
cluster properties of $R_n^{(2)}$ was solved
(at least, for a few of the simplest cluster algebras, and more generally conjectured)
in~\cite{Golden:2014xqa}.
The solution is a function associated to the $A_3$ cluster algebra
which we can write in the form
\begin{equation}
\label{eq:fA3def}
f_{A_3}(x_1,x_2,x_3) = \sum_{i=1}^3 K_{2,2}(x_{i,1}, x_{i,2}),
\end{equation}
where
\begin{alignat}{2}\label{eq:A3coords}
        x_{1,1} &= x_1,  &\quad x_{1,2} &= 1/x_3,\nonumber \\
        x_{2,1} &= \left(x_1 x_2+x_2+1\right) x_3, & x_{2,2}
        &=\frac{x_1 x_2+x_2+1}{x_1}, \\
    x_{3,1} &= \frac{x_2 x_3+x_3+1}{x_2}, & x_{3,2} &= \frac{x_2 x_3+x_3+1}{x_1 x_2 x_3} \nonumber
\end{alignat}
and
\begin{equation}
K_{2,2}(x,y) = L_{2,2}(x,y) - \left[
\Li_4(x/y) - \frac{1}{3} \Li_3(x/y) \log(x/y) - (x \leftrightarrow y)\right] - \frac{1}{2} \Li_2(-x)
\Li_2(-y).
\end{equation}
The expression for $K_{2,2}$ given here differs from the one presented
in~\cite{Golden:2014xqa} by the addition of terms proportional to products
of logarithms as well as the final $\Li_2 \Li_2$ term, none of which affect the coproduct
of $K_{2,2}$.

As long as the three $x_i$ generate an $A_3$ algebra
$x_1 \to x_2 \to x_3$ (which could be a subalgebra of a larger
algebra), the $A_3$ function accomplishes a remarkable feat:
\begin{itemize}
\item the $\Lambda^2 B_2$ component of its coproduct, $\sum_{i=1}^3
\{x_{i,1}\}_2 \wedge \{x_{i,2}\}_2$, involves only pairs of Poisson commuting
$\mathcal{X}$-coordinates;
\item the $B_3 \otimes \mathbb{C}^*$ component of its coproduct can be written
in terms of $\mathcal{X}$-coordinates (the $\Li_4$ term in $K_{2,2}$ is
crucial here);
\item its symbol can be written entirely in terms of $\mathcal{A}$-coordinates
(here the $\Li_3 \log$ term is crucial);
\item and it is smooth and real-valued
everywhere inside the positive domain
(i.e., as long as $x_1,x_2,x_3>0$), thanks to the terms which were
added compared to~\cite{Golden:2014xqa}.
\end{itemize}
The $\Li_2 \Li_2$ term in~(\ref{eq:fA3def}) is completely innocuous and was chosen
for inclusion because it was observed to nicely package together most of the
$\Li_2 \Li_2$ terms in the amplitude $R_7^{(2)}$.
It would be very interesting to see if a more optimal packaging of subleading terms could
be obtained, whether for $n=7$ or even for all $n$.

Working with $A_3$ functions, rather than the underlying individual
$L_{2,2}$'s, therefore allows us to avoid having to enlarge the symbol
alphabet beyond the set of cluster $\mathcal{A}$-coordinates.  Moreover,
when expressing the classical contributions to an amplitude we are
able to restrict our attention to the functions $\Li_k(-x)$, which are
smooth and real-valued throughout the positive domain as long
as the arguments $x$ are always taken from the set of cluster
$\mathcal{X}$-coordinates.

\section{The Algorithm for $R_n^{(2)}$}

The algorithm is naturally broken into four steps.
(1) As discussed in the previous section,
we start by
writing down a linear combination of $A_3$ cluster functions
with the same $\Lambda^2 B_2$
content
as the desired $R_n^{(2)}$.
After subtracting this linear
combination from the amplitude we are left with a function
which (2) we express in terms of the classical polylogarithms $\Li_k$
following the algorithm described in~\cite{Goncharov:2010jf}.
One
minor difference with respect to~\cite{Goncharov:2010jf} is that
we prioritize the $\Li_4$ terms over those which can be written as
products of lower-weight $\Li_k$'s, since only the former contribute to
the $B_3 \otimes \mathbb{C}^*$ component of the coproduct.
So, to be explicit, we proceed in the following order:
$f_{A_3}$, $\Li_4$, $\Li_2 \Li_2$, $\Li_2 \log \log$, $\Li_3 \log$,
$\log \log \log \log$.

At this stage we have a function with the same symbol as the amplitude,
so the difference is expected to be equal to $\pi^2$ times polylogarithm
functions of weight two.  We ought not find
any terms proportional to $i \pi$ times a function of weight three since
at each step we work with functions that are manifestly free of branch
cuts in the positive domain.
(3) The $\mathcal{O}(\pi^2)$ terms can be found by comparison
to the known~\cite{CaronHuot:2011ky,Golden:2013lha}
all-$n$ formula for the differential $dR_n^{(2)}$ of the amplitude.
(4) Finally, the overall additive constant in the amplitude can be determined
by enforcing smoothness of the collinear limit $R_n^{(2)} \to
R_{n-1}^{(2)}$, a property which is built
into the definition of the remainder function~\cite{Bern:2005iz}.

\section{The Construction of $R_7^{(2)}$}

We present here some details about the expression for $R_7^{(2)}$ generated
by our algorithm.  Some of the contributions, in particular the
terms of the form $\Li_2 \log \log$ or $\log \log \log \log$, are too
numerous to reasonably display in the text, so we refer the reader
to the Mathematica file associated to this note for the full symbolic
result\footnote{In case of any discrepancy between formulas in the
text and the Mathematica file, the latter is authoritative.}.

We begin by recalling the representation of the non-classical pieces
of $R_7^{(2)}$ in terms of $A_3$ functions, presented
in~\cite{Golden:2014xqa} as
\begin{multline}\label{eq:fA3sum}
\frac{1}{2} f_{A_3}\left(\textstyle{\frac{\langle 1245\rangle  \langle
   1567\rangle }{\langle 1257\rangle  \langle 1456\rangle }},
   \frac{\langle 1235\rangle  \langle 1456\rangle }{\langle
   1256\rangle  \langle 1345\rangle }, \frac{\langle 1234\rangle
   \langle 1257\rangle }{\langle 1237\rangle  \langle 1245\rangle
   }\right)+\frac{1}{2} f_{A_3}\left(\textstyle{\frac{\langle 1345\rangle  \langle
   1567\rangle }{\langle 1357\rangle  \langle 1456\rangle }},
   \frac{\langle 1235\rangle  \langle 3456\rangle }{\langle
   1356\rangle  \langle 2345\rangle }, \frac{\langle 1234\rangle
   \langle 1357\rangle }{\langle 1237\rangle  \langle 1345\rangle
   }\right)\\+\text{ dihedral} +\text{parity~conjugate.}
\end{multline}
As we emphasized in~\cite{Golden:2014xqa},
the difference between $R^{(2)}_7$ and (\ref{eq:fA3sum}) is a weight-four
polynomial in the functions $\Li_k(-x)$ for $k=1,2,3$ (and $\pi^2$), with arguments $x$ drawn from the
385 $\mathcal{X}$-coordinates of the $\Gr(4,7)$ cluster algebra.

The $B_3 \otimes \mathbb{C}^*$ component of the coproduct
of $R_7^{(2)}$ was computed in~\cite{Golden:2013xva}.  We find that
the $\Li_4$ terms
\begin{multline}
	-\text{Li}_4\left(\textstyle{-\frac{\langle 1234\rangle  \langle 1256\rangle
   }{\langle 1236\rangle  \langle 1245\rangle
   }}\right)-\text{Li}_4\left(\textstyle{-\frac{\langle 1234\rangle  \langle
   1257\rangle }{\langle 1237\rangle  \langle 1245\rangle
   }}\right)-\frac{1}{2} \text{Li}_4\left(\textstyle{-\frac{\langle 1234\rangle
    \langle 1357\rangle }{\langle 1237\rangle  \langle 1345\rangle
   }}\right)-\frac{1}{2} \text{Li}_4\left(\textstyle{-\frac{\langle 1234\rangle
    \langle 1456\rangle }{\langle 1246\rangle  \langle 1345\rangle
   }}\right)\\+\text{ dihedral} +\text{parity~conjugate,}
\label{eq:li4}
\end{multline}
must be added to eq.~(\ref{eq:fA3sum}) in order to correctly reproduce
the full coproduct of the amplitude.

At this stage we know that the difference between $R_7^{(2)}$
and eqs.~(\ref{eq:fA3sum}) plus~(\ref{eq:li4})
is a product of $\Li_k$ functions of weight strictly less than four.
Following the procedure outlined in~\cite{Goncharov:2010jf} we find that
the missing $\Li_2 \Li_2$ terms (beyond the ones that we have already snuck
in via eq.~(\ref{eq:fA3def})) are
\begin{multline}
	\text{Li}_2\left(\textstyle{\frac{\langle 3(17)(24)(56)\rangle }{\langle
   1237\rangle  \langle 3456\rangle }}\right)
   \text{Li}_2\left(\textstyle{\frac{\langle 1456\rangle  \langle
   3(17)(24)(56)\rangle }{\langle 1234\rangle  \langle 1567\rangle
   \langle 3456\rangle }}\right)+\text{Li}_2\left(\textstyle{{-}\frac{\langle
   1234\rangle  \langle 1257\rangle }{\langle 1237\rangle  \langle
   1245\rangle }}\right) \text{Li}_2\left(\textstyle{{-}\frac{\langle 1234\rangle
   \langle 1457\rangle }{\langle 1247\rangle  \langle 1345\rangle
   }}\right)\\-\frac{1}{2} \text{Li}_2\left(\textstyle{{-}\frac{\langle 1234\rangle
   \langle 1257\rangle }{\langle 1237\rangle  \langle 1245\rangle
   }}\right) \text{Li}_2\left(\textstyle{{-}\frac{\langle 1245\rangle  \langle
   1567\rangle }{\langle 1257\rangle  \langle 1456\rangle
   }}\right)-\text{Li}_2\left(\textstyle{{-}\frac{\langle 1234\rangle  \langle
   1357\rangle }{\langle 1237\rangle  \langle 1345\rangle }}\right)
   \text{Li}_2\left(\textstyle{{-}\frac{\langle 1345\rangle  \langle 1567\rangle
   }{\langle 1357\rangle  \langle 1456\rangle
   }}\right)\\-\text{Li}_2\left(\textstyle{{-}\frac{\langle 1237\rangle  \langle
   1467\rangle }{\langle 1267\rangle  \langle 1347\rangle }}\right)
   \text{Li}_2\left(\textstyle{{-}\frac{\langle 1236\rangle  \langle 2567\rangle
   }{\langle 1267\rangle  \langle 2356\rangle
   }}\right)+\text{Li}_2\left(\textstyle{{-}\frac{\langle 1236\rangle  \langle
   2567\rangle }{\langle 1267\rangle  \langle 2356\rangle }}\right)
   \text{Li}_2\left(\textstyle{{-}\frac{\langle 2345\rangle  \langle 3467\rangle
   }{\langle 2347\rangle  \langle 3456\rangle }}\right)\\+\text{ dihedral} +\text{parity~conjugate.}
\label{eq:li2li2}
\end{multline}
We also find the $\Li_3 \log$ terms
\begin{multline}
	\left(\frac{1}{2} \text{Li}_3\left(\textstyle{{-}\frac{\langle 1267\rangle
   \langle 1456\rangle }{\langle 1246\rangle  \langle 1567\rangle
   }}\right)-\frac{1}{2} \text{Li}_3\left(\textstyle{{-}\frac{\langle 1246\rangle
   \langle 1345\rangle }{\langle 1234\rangle  \langle 1456\rangle
   }}\right)\right) \log \left(\textstyle{\frac{\langle 1237\rangle  \langle
   1246\rangle }{\langle 1234\rangle  \langle 1267\rangle
   }}\right)\\+\Bigg(-\frac{1}{2} \text{Li}_3\left(\textstyle{{-}\frac{\langle
   1237\rangle  \langle 1345\rangle }{\langle 1234\rangle  \langle
   1357\rangle }}\right)-\frac{1}{2} \text{Li}_3\left(\textstyle{{-}\frac{\langle
   1247\rangle  \langle 1345\rangle }{\langle 1234\rangle  \langle
   1457\rangle }}\right)+\text{Li}_3\left(\textstyle{{-}\frac{\langle 1257\rangle
   \langle 1347\rangle }{\langle 1237\rangle  \langle 1457\rangle
   }}\right)+\text{Li}_3\left(\textstyle{{-}\frac{\langle 1257\rangle  \langle
   1456\rangle }{\langle 1245\rangle  \langle 1567\rangle
   }}\right)\\+\frac{1}{2} \text{Li}_3\left(\textstyle{{-}\frac{\langle 1267\rangle
   \langle 1456\rangle }{\langle 1246\rangle  \langle 1567\rangle
   }}\right)+\frac{1}{2} \text{Li}_3\left(\textstyle{{-}\frac{\langle 1357\rangle
   \langle 1456\rangle }{\langle 1345\rangle  \langle 1567\rangle
   }}\right)-\text{Li}_3\left(\textstyle{{-}\frac{\langle 1235\rangle  \langle
   1267\rangle  \langle 1457\rangle }{\langle 1237\rangle  \langle
   1245\rangle  \langle 1567\rangle }}\right)\Bigg) \log
   \left(\textstyle{\frac{\langle 1247\rangle  \langle 1345\rangle }{\langle
   1234\rangle  \langle 1457\rangle
   }}\right)\\+\Bigg(-\text{Li}_3\left(\textstyle{{-}\frac{\langle 1236\rangle
   \langle 1245\rangle }{\langle 1234\rangle  \langle 1256\rangle
   }}\right)-\frac{1}{2} \text{Li}_3\left(\textstyle{{-}\frac{\langle 1237\rangle
   \langle 1245\rangle }{\langle 1234\rangle  \langle 1257\rangle
   }}\right)+\text{Li}_3\left(\textstyle{{-}\frac{\langle 1247\rangle  \langle
   1256\rangle }{\langle 1245\rangle  \langle 1267\rangle
   }}\right)+\frac{1}{2} \text{Li}_3\left(\textstyle{{-}\frac{\langle 1237\rangle
   \langle 1345\rangle }{\langle 1234\rangle  \langle 1357\rangle
   }}\right)\\+\frac{1}{2} \text{Li}_3\left(\textstyle{{-}\frac{\langle 1246\rangle
   \langle 1345\rangle }{\langle 1234\rangle  \langle 1456\rangle
   }}\right)+\text{Li}_3\left(\textstyle{{-}\frac{\langle 1247\rangle  \langle
   1345\rangle }{\langle 1234\rangle  \langle 1457\rangle
   }}\right)-\frac{1}{2} \text{Li}_3\left(\textstyle{{-}\frac{\langle 1257\rangle
   \langle 1456\rangle }{\langle 1245\rangle  \langle 1567\rangle
   }}\right)-\frac{1}{2} \text{Li}_3\left(\textstyle{{-}\frac{\langle 1267\rangle
   \langle 1456\rangle }{\langle 1246\rangle  \langle 1567\rangle
   }}\right)\\-\frac{1}{2} \text{Li}_3\left(\textstyle{{-}\frac{\langle 1456\rangle
   \langle 2345\rangle }{\langle 1245\rangle  \langle 3456\rangle
   }}\right)-\text{Li}_3\left(\textstyle{{-}\frac{\langle 1457\rangle  \langle
   2345\rangle }{\langle 1245\rangle  \langle 3457\rangle
   }}\right)+\text{Li}_3\left(\textstyle{{-}\frac{\langle 1457\rangle  \langle
   2456\rangle }{\langle 1245\rangle  \langle 4567\rangle
   }}\right)\Bigg) \log \left(\textstyle{\frac{\langle 1237\rangle  \langle
   1245\rangle }{\langle 1234\rangle  \langle 1257\rangle }}\right)\\+\text{ dihedral} +\text{parity~conjugate.}
\label{eq:li3log}
\end{multline}
The remaining $\Li_2 \log \log$ and
$\log\log\log\log$
terms which must be added to eqs.~(\ref{eq:fA3sum}), (\ref{eq:li4}), (\ref{eq:li2li2}) and~(\ref{eq:li3log})
in order to fully match the known symbol of $R_7^{(2)}$ are too numerous to display here
and are recorded in the attached Mathematica file.

Next we turn to the problem of fixing ``beyond-the-symbol'' terms, given by numerical constants (in this
application, rational numbers times $\pi^k$) times functions of weight $4-k$.
The terms proportional to $\pi^2$
may be deduced by computing the full differential of all of the terms we have accumulated so
far, and subtracting the result from the known analytic formula for
$dR_7^{(2)}$~\cite{CaronHuot:2011ky,Golden:2013lha}.  The result is a linear combination (with rational coefficients)
of terms like $\pi^2 \log(a_1) d \log a_2$ for various $\mathcal{A}$-coordinates $a_1,a_2$.
This can be integrated analytically to a linear combination of terms like $\pi^2 \Li_2(-x_i)$ and
$\pi^2 \log(x_j) \log(x_k)$ with all arguments being $\mathcal{X}$-coordinates.
In this manner we find that the $\pi^2 \Li_2$ terms in our representation of $R_7^{(2)}$ are given by
\begin{equation}
\frac{7\pi^2}{48} \Li_2\left(\textstyle{-\frac{\ket{1247}\ket{1345}}{\ket{1234}\ket{1457}}}\right)
-\frac{\pi^2}{8} \Li_2\left(\textstyle{-\frac{\ket{1(23)(45)(67)}}{\ket{1234}\ket{1567}}}\right)
\\+\text{dihedral} +\text{parity~conjugate,}
\end{equation}
while the $\pi^2 \log \log$ terms are again somewhat too numerous to efficiently display here.

At this point we have constructed a function which agrees with $R_7^{(2)}$ up to a single
overall additive constant\footnote{There are no  $\zeta(3) \log$ terms
since $dR_n^{(2)}$ is known~\cite{CaronHuot:2011ky}
to not contain any terms proportional
to $\zeta(3)$.}.
This constant, expected to be a rational number times $\pi^4$,
can be determined
by the requirement that $R_7^{(2)} \to R_6^{(2)}$ smoothly
in the collinear limit.
We choose to parameterize the $6 \parallel 7$ collinear limit
following~\cite{CaronHuot:2011ky} by replacing
\begin{equation}
Z_7 \to Z_7(t) = Z_6 - t (\alpha Z_1 + \beta Z_5) + t^2 Z_2,
\end{equation}
with $\alpha$ and $\beta$ being arbitrary parameters, and then taking
the limit $t \to 0$.  As long as the
starting point $(Z_1,\ldots,Z_7)$ is inside the positive domain
and
$\alpha$ and $\beta$ are chosen to be positive,
then there exists a finite $t_0 > 0$
such that
$(Z_1,\ldots,Z_6,Z_7(t))$
lies in the positive domain for all $0 < t < t_0$.
Then the collinear limit\footnote{We caution the reader that
our normalization convention for $R_7^{(2)}$ agrees with that
of~\cite{CaronHuot:2011ky}, which differs by a factor of four
from that of~\cite{Goncharov:2010jf}, so the $R_6^{(2)}$ appearing
on the right-hand side of eq.~(\ref{eq:collinear}) should be four times
the function $R_6^{(2)}$ given in the latter reference.}
\begin{equation}
\label{eq:collinear}
\lim_{t \to 0^+} R^{(2)}_7(Z_1,\ldots,Z_6,Z_7(t)) = R^{(2)}_6(Z_1,\ldots,Z_6),
\end{equation}
together with the known formula~\cite{Goncharov:2010jf}
for $R_6^{(2)}$, determines the overall additive constant in $R_7^{(2)}$.

Each cross-ratio appearing our formula for $R_7^{(2)}$ approaches either
$0$, $\infty$, or a finite value in the limit $t \to 0^+$, so
it is a simple matter to compute the limit of the formula using the asymptotic
behavior of the polylogarithm functions
\begin{align}
\Li_2(-1/t) &\sim - \frac{1}{2} \log^2 t - \frac{\pi^2}{2},
\\
\Li_3(-1/t) &\sim + \frac{1}{6} \log^3 t + \frac{\pi^2}{6} \log t,
\\
\Li_4(-1/t) &\sim - \frac{1}{24} \log^4 t - \frac{\pi^2}{12} \log^2 t
- \frac{7 \pi^4}{360}
\end{align}
together with the asymptotic expansions
(when $x$, $t$ and $a$ are positive)
\begin{align}
L_{2,2}(x,t) &\sim 0,
\\
L_{2,2}(x,1/t) &\sim \frac{1}{4} \Li_2(-x) \log^2t + \Li_3(-x) \log t
+ \Li_4(-x) + \frac{\pi^2}{12} \Li_2(-x),
\\
L_{2,2}(1/t,a/t^2) &\sim - \frac{5}{24} \log^4 t + \frac{1}{3} \log a \log^3 t
- \frac{1}{8} \log^2 a \log^2 t + \frac{\pi^4}{24} \log^2 t - \frac{\pi^2}{24} \log^2 a
- \frac{\pi^4}{30},
\end{align}
where $\sim$ signifies the omission of terms which vanish as powers of $t$
(or powers of $t$ times powers of $\log t$).
We have taken the limit of $R_7^{(2)}$
by choosing various random initial kinematic points in the positive
domain with all
momentum twistors having integer entries.
Then, after taking the limit $t \to 0^+$, the two sides of
eq.~(\ref{eq:collinear}) can be evaluated numerically
with arbitrary precision.
In this manner we find that that we have to add $- \frac{13}{36} \pi^4$ to
our formula for $R_7^{(2)}$
in order for eq.~(\ref{eq:collinear})
to be satisfied.

\section{The Function $R_7^{(2)}$}

Several very different ingredients have gone into the construction of our formula
for $R_7^{(2)}$, from Caron-Huot's calculation of symbols via an extension of
superspace to the mathematical structure of cluster algebras.
As an independent test that all of these ingredients have been put together correctly
it is reassuring to compare our result to numerical values for $R_7^{(2)}$
obtained
in~\cite{Anastasiou:2009kna} via the
Wilson loop approach to scattering amplitudes in SYM theory.

To get an intuition for a function it is often useful to see a plot of it, such as
fig.~11 of~\cite{Anastasiou:2009kna} which shows $R_7^{(2)}$ evaluated on the
``symmetric line'', the locus where
\begin{equation}
(u_{14}, u_{25}, u_{36}, u_{47}, u_{15}, u_{26}, u_{37}) = (u,u,u,u,u,u,u)
\end{equation}
in terms of
\begin{equation}
u_{ij} = \frac{\ket{i\,i{+}1\,j{+}1\,j{+}2}
\ket{i{+}1\,i{+}2\,j\,j{+}1}}{\ket{i\,i{+}1\,j\,j{+}1} \ket{i{+}1\,i{+}2\,j{+}1\,j{+}2}}.
\end{equation}

When the seven momentum vectors of the scattering particles are required to lie
in four spacetime dimensions, the $u_{ij}$ are not free (indeed they cannot be,
since the dimension of $\Conf_7(\mathbb{P}^3)$ is only six) but are constrained
to satisfy a particular seventh-order polynomial equation called the Gram determinant
constraint.
The symmetric line intersects the Gram locus only at isolated points
(specifically, at the roots of
$(1+u)(1-4u+3 u^2+u^3)^2)$.
The authors of~\cite{Anastasiou:2009kna} evaded this constraint by
allowing the momenta to lie in arbitrary dimension.  By making use of momentum
twistor machinery our result for $R_7^{(2)}$ is solidly tied to four-dimensional
kinematics, although we anticipate that it should not be very difficult
to relax this constraint.

Until that is done
we are therefore unable
to provide a plot of our $R_7^{(2)}$ formula along the symmetric line.
Instead we display in fig.~1 a plot of this function along the line
segment
\begin{equation}
\label{eq:line}
(u_{14}, u_{25}, u_{36}, u_{47}, u_{15}, u_{26}, u_{37}) = (u,u,u,u,u,u,
\textstyle{\frac{(1 - u - u^2)^2}{1 - 2 u^2}})
\end{equation}
which satisfies the Gram constraint for all $u$ and
which lies in the positive domain for $0 < u < u_0 = 0.35689586789\ldots$, this number
being the smallest positive root of $1-4 u + 3 u^2 + u^3 =0$.

The endpoint of this line segment at $u = u_0$ is rather
special\footnote{This point is a close analog to the special
point $(u_{14},u_{25},u_{36}) = (\frac{1}{4},\frac{1}{4},\frac{1}{4})$
in six-particle kinematics.}:
it touches the symmetric
line at the tip of the positive domain.  At this special point
we find
\begin{equation}
\label{eq:data1}
R^{(2)}_7(u_0,u_0,u_0,u_0,u_0,u_0,u_0) = 10.4368451968\ldots\,.
\end{equation}
At a conveniently chosen non-symmetric point point satisfying the Gram
constraint we find for example
\begin{equation}
\label{eq:data2}
R^{(2)}_7(\textstyle{\frac{1}{4}},\textstyle{\frac{1}{4}},\textstyle{\frac{1}{4}},\textstyle{\frac{1}{4}},\textstyle{\frac{1}{4}},\textstyle{\frac{1}{4}},\textstyle{\frac{121}{224}}) = 23.8717248322\ldots\,.
\end{equation}
Both of these values are consistent with numerical results
obtained using the Wilson loop computation of~\cite{Anastasiou:2009kna}\footnote{We
thank A.~Brandhuber, P.~Heslop and G.~Travaglini for
correspondence and for providing
us with their results at these kinematic points, which
match eqs.~(\ref{eq:data1}) and~(\ref{eq:data2}) to $0.003\%$, within
the estimated margin of error of their numerical calculation.}.
At all points in the positive
domain where we have evaluated $R_7^{(2)}$, we have always
found it take positive values, supporting the conjecture
of~\cite{positive}.

\begin{figure}
\begin{center}
\includegraphics[width=4.0in]{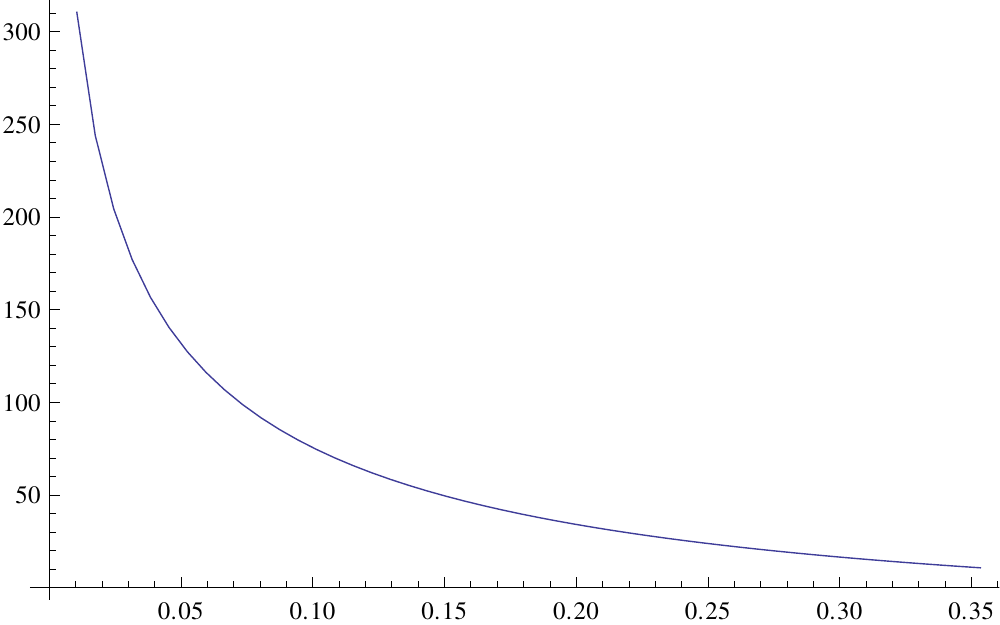}
\end{center}
\caption{The two-loop seven-point MHV remainder function $R_7^{(2)}$ evaluated
along the line segment parameterized by eq.~(\ref{eq:line}) between $u=0$ and
the boundary of the positive domain at $u\approx 0.3569$.}
\end{figure}

\section{Conclusion}

We have described an algorithm for bootstrapping
explicit analytic formulas for the two-loop $n$-point MHV remainder functions
$R_n^{(2)}$ in SYM theory
from known results in the literature
for the symbol~\cite{CaronHuot:2011ky} and the differential~\cite{Golden:2013lha} of $R_n^{(2)}$.
The algorithm expresses these amplitudes as linear combination of $A_3$ cluster polylogarithm
functions~\cite{Golden:2014xqa} and (products of) classical polylogarithm functions $\Li_k(-x)$
with arguments $x$ drawn from the set of $\mathcal{X}$-coordinates~\cite{FG03b} for
the $\Gr(4,n)$ cluster algebra.  Each building block utilized in the construction is
manifestly real-valued and singularity-free inside the positive domain, ensuring that the
generated formula
for $R_n^{(2)}$ manifests this property as well.

As a sample application of this algorithm we have constructed an explicit analytic
representation for $R_7^{(2)}$.
We would like to emphasize that we have put  almost no effort into optimizing our result, instead opting
to see what we get by treating this as nothing more than a ``shift-enter'' computation.
Although we were somewhat surprised that the answer came out as compact as it did,
we anticipate that our result for $R_7^{(2)}$ will not be the ``best'' representation
possible but hope that it may provide a useful starting point for further exploration of the
structure of this amplitude.  In that sense we suspect our representation for $R_7^{(2)}$
may be more similar to the DDS formula~\cite{DelDuca:2009au,DelDuca:2010zg}
than to the GSVV formula~\cite{Goncharov:2010jf} for $R_6^{(2)}$.

Let us end by speculating about some possible ways in which our representation for
$R_7^{(2)}$ (and $R_n^{(2)}$ more generally) ought to be improved.
As a general statement, it is our hope that amplitudes should admit natural functional
representations which are as canonical as possible\footnote{Given the various known, but classifiable,
polylogarithm identities---the $5$-term $A_2$ identity for $\Li_2$, the
40-term $D_4$ identity for $\Li_3$, and other possibly existent identities not yet discovered.}
and that any unexplained ambiguity in how to write an amplitude should be a cause for disappointment.
This is because
our ultimate dream
is that it should be possible to formulate a list of physical and
mathematical constraints which determine SYM theory amplitudes uniquely, and any free parameter
appearing in the representation of some amplitude represents a lost opportunity to manifest some
otherwise hidden property it satisfies.

For example, we find it suboptimal that (as mentioned in~\cite{Golden:2014xqa})
the non-classical part of $R_7^{(2)}$ may be expressed in many different ways in terms of $A_3$ functions.
It would be fantastic if one could identify some particular $A_3$ subalgebras inside the $\Gr(4,7)$ cluster
algebra (or $\Gr(4,n)$ more generally) which are for some reason preferred for expressing
two-loop MHV amplitudes.
Moreover it would be nice if all of the classical terms tabulated in section 4
could be absorbed
into an appropriate
redefinition of the $A_3$ function given in eq.~(\ref{eq:fA3def}) so that the complete formula
for $R_7^{(2)}$, or even all $R_n^{(2)}$,
could be written as a simple linear combination of suitably
defined $A_3$ functions and nothing else.
If this magic $A_3$ function were positive-valued inside the positive domain,
it would furthermore manifest the conjectured~\cite{positive}
positivity of $R_n^{(2)}$ itself.
It would be ideal if this could be done for all $n$ in a way which
manifests collinear limits, with the various $A_3$ functions appearing in $R_n^{(2)}$ tending either to
zero or to $n{-}1$-point $A_3$ functions in the collinear limit.
Finally, perhaps it is not the $A_3$ function but something else which
is the right building block for realizing all of these dreams.

\acknowledgments

We thank L.~Dixon
and G.~Papathanasiou for comments on the draft, and
acknowledge having benefitted
from stimulating discussions and longstanding collaboration on related problems with
M.~Paulos,
C.~Vergu, A.~Volovich, and especially A.~Goncharov.
This work was supported by the US Department of Energy under contract
DE-SC0010010 Task A.

\end{document}